\documentclass[prl,twocolumn,showpacs,preprintnumbers,nofootinbib]{revtex4}

\usepackage{amsmath}
\usepackage{graphicx}
\usepackage{dcolumn}
\usepackage{epsfig}
\usepackage{verbatim}
\usepackage{psfrag}
\usepackage{bm}
\usepackage{bbm}
\usepackage{latexsym}
\usepackage[english]{babel}
\usepackage{amsthm}
\usepackage{color}
\usepackage{amssymb}

\begin{document}

\preprint{ITP-UU-11/16, SPIN-11/10}

\title{The Cosmological Constant and Lorentz Invariance of the Vacuum State}

\author{Jurjen F. Koksma\footnote{J.F.Koksma@uu.nl, T.Prokopec@uu.nl}}
\affiliation{Institute for Theoretical
Physics (ITP) \& Spinoza Institute, Utrecht University, Postbus
80195, 3508 TD Utrecht, The Netherlands}

\author{Tomislav Prokopec$^*$}
\affiliation{Institute for Theoretical Physics (ITP) \& Spinoza
Institute, Utrecht University, Postbus 80195, 3508 TD Utrecht, The
Netherlands}

\begin{abstract}
One hope to solve the cosmological constant problem is to identify a symmetry principle, based on which the cosmological constant can be reduced either to zero, or to a tiny value. Here, we note that requiring that the vacuum state is Lorentz invariant significantly reduces the theoretical value of the vacuum energy density. Hence, this also reduces the discrepancy between the observed value of the cosmological constant and its theoretical expectation, down from 123 orders of magnitude to 56 orders of magnitude. We find that, at one loop level, massless particles do not yield any contribution to the cosmological constant. Another important consequence of Lorentz symmetry is stabilization of the gravitational hierarchy: the cosmological constant (divided by Newton's constant) does not run as the quartic power of the renormalization group scale, but instead only logarithmically.
\end{abstract}


\maketitle

\textsc{Introduction} --- The Universe's recently observed accelerated expansion is in standard $\Lambda$CDM cosmology modeled by a cosmological constant $\Lambda_{\mathrm{eff}}$ in the Einstein Field equations~\cite{Weinberg:1988cp, Peebles:2002gy, Nobbenhuis:2004wn, Silvestri:2009hh}, which in the semiclassical approach to quantum gravity read:
\begin{equation}\label{EinsteinQFE}
R_{\mu\nu}-\frac{1}{2}R g_{\mu\nu} +\Lambda g_{\mu\nu} = 8\pi G
\langle \Omega | \hat{T}_{\mu\nu}|\Omega\rangle\,.
\end{equation}
Here, $\langle  \Omega | \hat{T}_{\mu\nu}| \Omega  \rangle= T_{\mu\nu}+
\langle  \Omega | \hat{\delta T}_{\mu\nu}| \Omega \rangle $ consists of both classical and quantum contributions, and $|\Omega\rangle$ is a quantum state. Let us for simplicity's sake consider one free scalar field $\phi$ in a curved spacetime, whose stress-energy tensor is given by:
\begin{equation}\label{stressenergytensor}
\hat{T}_{\mu\nu} = \partial_\mu \hat{\phi}
\partial_\nu \hat{\phi} - g_{\mu\nu}\! \left[\frac{1}{2}\partial_{\alpha}\hat{\phi}(x)
\partial_{\beta} \hat{\phi}(x) g^{\alpha\beta} \! +
\frac{1}{2}m^{2}\hat{\phi}^{2}(x)\right]\!.
\end{equation}
where $g_{\mu\nu}$ is the metric tensor. Here we are primarily interested in the flat spacetime limit, in which $g_{\mu\nu}$ reduces to a flat Minkowski metric, $g_{\mu\nu}={\rm diag} (-1,1,1,1)$. In this case, the energy density per mode for a scalar field in its ground state equals $\omega/2$, where $\omega^2=k^2+m^2$, and $k=\|\vec{k}\|$ as usual. The total vacuum energy density is therefore quartically divergent in the ultraviolet (UV):
\begin{equation}\label{vacuumenergy}
\langle 0| \hat{\delta T}_{00}^{\mathrm{vac}} |0 \rangle = \langle
0| \hat{\rho}_{\mathrm{vac}} |0 \rangle = \int
\frac{\mathrm{d}^3\vec{k}}{(2\pi)^3} \frac{1}{2} \sqrt{k^2+m^2}
\,.
\end{equation}
From observations~\cite{Will:2005va} we know to a high accuracy that the (quantum) vacuum is Lorentz invariant, and Lorentz symmetry is therefore, like unitarity \cite{'tHooft:1973pz} and causality \cite{Koksma:2010zy, Westra:2010zx}, a crucial aspect of any properly formulated quantum field theory. Hence, this vacuum energy density must give rise to a stress-energy tensor in Minkowski spacetime of the form:
\begin{equation}\label{vacuumstressenergytensor}
\langle 0| \hat{\delta T}_{\mu\nu}^{\mathrm{vac}} |0 \rangle = -
\langle 0| \hat{\rho}_{\mathrm{vac}} |0 \rangle g_{\mu\nu}
\,.
\end{equation}
A perfect fluid form would apply in homogeneous and isotropic FLRW or Friedmann-Lema\^itre-Robertson-Walker spacetimes. We can combine the geometric and matter contributions to the cosmological constant in an effective cosmological constant:
\begin{equation}\label{effectivecosmologicalconstant}
\Lambda_{\mathrm{eff}} = \Lambda + 8\pi G  \langle 0|
\hat{\rho}_{\mathrm{vac}} |0 \rangle  \,,
\end{equation}
or, equivalently, in an effective energy density of the vacuum:
\begin{equation}\label{effectivecosmologicalconstant2}
\rho_{\mathrm{eff}} = \frac{\Lambda}{8\pi G} +  \langle 0|
\hat{\rho}_{\mathrm{vac}}|0 \rangle  \,.
\end{equation}
Let us examine these statements in a little more detail. As the expression in equation~(\ref{vacuumenergy}) is
divergent, one can formally introduce a UV cutoff $\Lambda_{\mathrm{UV}}$ and evaluate the integral in equation~(\ref{vacuumenergy}):
\begin{align}\label{vacuumenergy2}
& \langle 0| \hat{\rho}_{\mathrm{vac}} |0 \rangle =
\int_{0}^{\Lambda_{\mathrm{UV}}} \frac{\mathrm{d}k \, k^2}{4 \pi^2}\sqrt{k^2+m^2}\\
&=\frac{\Lambda_{\mathrm{UV}}^4}{16\pi^2} +
\frac{m^2\Lambda_{\mathrm{UV}}^2}{16\pi^2} +
\frac{m^4}{64\pi^2}\log\left[\frac{m^2 e^{1/2}}{4 \Lambda_{\mathrm{UV}}^2}\right]\! +\!
\mathcal{O}\left(\Lambda_{\mathrm{UV}}^{-1}\right) \nonumber ,
\end{align}
where we evaluate the integral on the first line exactly and consider its leading order behaviour only. As we will show shortly, the first two terms in this expression break Lorentz invariance and we would therefore consider these terms to be unphysical. In most of the literature, however, the cutoff $\Lambda_{\mathrm{UV}}$ introduced in equation~(\ref{vacuumenergy2}) is considered to be physical and is not removed in order to obtain a renormalised expression for the cosmological constant. The leading order term proportional to $\Lambda_{\mathrm{UV}}^4$ is thus kept, and it is then argued that perhaps the Planck scale is a natural cutoff, as that scale roughly corresponds to the scale up to which we trust perturbative calculations in quantum field theory on curved spacetimes. Then, one finds:
\begin{equation}\label{vacuumenergy3}
\langle 0| \hat{\rho}_{\mathrm{vac}} |0 \rangle \sim \Lambda_{\mathrm{UV}}^4 \sim
M_{\mathrm{pl}}^4 \sim 10^{76} \, \mathrm{GeV}^4 \,,
\end{equation}
while the observed effective energy density of the vacuum is roughly given by:
\begin{equation}\label{effectivecosmologicalconstant3}
\rho_{\mathrm{eff}} \sim 10^{-47} \, \mathrm{GeV}^4 \,,
\end{equation}
where we assume that this energy density stems from a dark energy contribution. In order to accommodate the observed value above, the geometric contribution $\Lambda$ to $\rho_{\mathrm{eff}}$ has to be fine-tuned to 123 decimal places. The ``old'' cosmological constant problem can be phrased as follows: what is the origin of this extremely tiny energy density? The huge number in equation~(\ref{vacuumenergy3}) thus assumes that the cutoff $\Lambda_{\mathrm{UV}}$ is kept finite and should not be sent to infinity. However, it is precisely the latter operation one needs to perform in order to obtain a correct physical result that is both Lorentz invariant and cutoff independent, as we show in this paper.

There are other examples in nature in which symmetry is invoked to get rid of leading ultraviolet divergences. An important example is the transverse gluon mass, which in non-covariant gauges should naively acquire a contribution proportional to the cutoff. Due to gauge symmetry however, such contributions cancel. A second interesting example is chiral symmetry, which protects light quarks from acquiring large radiative masses.
We thus employ the symmetry principle of Lorentz invariance to reduce the theoretical value of the cosmological constant and stabilise the gravitational hierarchy.

\textsc{Cutoff Renormalisation} --- Let us evaluate the pressure from equation~(\ref{stressenergytensor}), arising from the vacuum contribution only, as:
\begin{align}\label{vacuumenergy4}
& \langle 0| \hat{P}_{\mathrm{vac}} |0 \rangle = \frac{1}{3}
\int^{\Lambda_{\mathrm{UV}}}_0 \frac{\mathrm{d}k \, k^2}{4 \pi^2}
\frac{k^2}{(k^2+m^2)^{\frac{1}{2}}} \\
&=\frac{\Lambda_{\mathrm{UV}}^4}{48\pi^2} - \frac{m^2
\Lambda_{\mathrm{UV}}^2}{48\pi^2} -
\frac{m^4}{64\pi^2}\log\left[\frac{m^2 e^{7/6}}{4 \Lambda_{\mathrm{UV}}^2}\right] \!+\!
\mathcal{O}\left(\Lambda_{\mathrm{UV}}^{-1}\right) \nonumber\,.
\end{align}
The divergences in the stress-energy tensor have to be regulated using for example a UV cutoff, dimensional, or Pauli-Villars regularisation. Let us here consider the first regularisation procedure, and discuss the other two methods in appendices A and B of this paper. We stress that the other two regularisation procedures give an identical final expression (\ref{vacuumenergy5}) for the renormalised stress-energy tensor. The first two terms proportional to $\Lambda_{\mathrm{UV}}^4$ and $m^2 \Lambda_{\mathrm{UV}}^2$ in both of the equations~(\ref{vacuumenergy2}) and~(\ref{vacuumenergy4}), which evidently break Lorentz invariance, can be removed by local counterterms. One can introduce a renormalisation scale $\mu$ to subtract the logarithmic divergences\footnote{This last step is not necessary to get a Lorentz invariant answer, but it is not essential for our argument. Subtracting the cutoff independent terms proportional to $m^4$ is necessary, however, as they break Lorentz symmetry.}. The final, physical result is obtained by sending $\Lambda_{\mathrm{UV}} \rightarrow \infty$, and absorbing the remaining numerical factors in the logarithm in a convenient manner, to find:
\begin{equation}\label{vacuumenergy5}
\langle 0| \hat{\rho}_{\mathrm{vac}}^{\mathrm{ren}} |0 \rangle  =
- \langle 0| \hat{P}_{\mathrm{vac}}^{\mathrm{ren}} |0 \rangle  =
\frac{m^4}{64 \pi^2}\log\left[\frac{m^2}{\mu^2}\right] \,.
\end{equation}
This indeed confirms equation~(\ref{vacuumstressenergytensor}). Provided that a naive regularisation scheme does not respect Lorentz symmetry, such as equation (\ref{vacuumenergy2}), we thus propose to modify the procedures of regularisation and renormalisation to respect Lorentz symmetry, resulting in a Lorentz invariant renormalisation (LIR) procedure. It is important to note that the logarithmic term containing the mass is considered to be physical unlike the other terms proportional to a power of the mass. The reason is that in a Higgs-like setting, where the mass $m$ of the field is generated by some other field, the logarithmic term cannot be subtracted by a local counterterm. We reproduce the result (\ref{vacuumenergy5}) in the appendices by making use of dimensional regularisation and Pauli-Villars regularisation.

\textsc{Lorentz Invariant Vacuum State} --- The vacuum state of a quantum theory in Minkowski spacetime is Lorentz invariant. Hence, its associated stress-energy tensor can only be of the form~(\ref{vacuumstressenergytensor}). We show in equation~(\ref{vacuumenergy5}) that this can only be achieved by removing the most divergent terms using a cutoff regularisation procedure. In literature, this simple observation seems to have been missed.

A second important observation, based on equation~(\ref{vacuumenergy5}) or~(\ref{vacuumenergy10}), is that $\Lambda_{\mathrm{eff}}/G$ runs logarithmically with the renormalisation scale $\mu$ (with a negative $\beta$-function for bosons, and a positive one for fermions). This stabilises the hierarchy of the cosmological constant, implying that if $\Lambda_{\mathrm{eff}}/G$ is small with respect to $m_{\mathrm{p}}^4$ for some scale $\mu$, it remains small for some other scale $\mu'$ (provided of course that $\mu$ is an observable scale).

Thirdly, note that, at one loop level, massless particles such as the photon and the graviton yield a vanishing contribution to the cosmological constant. More generally, at $n$ loops, we expect contributions proportional to $\lambda^{n-1}\Lambda^4$ and, in the massive case, $\lambda^{n-1} m^2\Lambda^2$, where $\lambda$ is the dimensionless quartic coupling constant. We expect that these terms break Lorentz invariance, too, such that they should be subtracted accordingly. In curved spacetimes, we expect, at one loop level, contributions proportional to $R \Lambda^2$, where $R$ is the Ricci scalar. In cosmology, $R \sim H^2$, where $H(t)$ is the Hubble parameter, such that this term does not contribute to the cosmological term in any spacetime other than de Sitter spacetime where $H$ is a global constant.

Let us now estimate the change of the value of the matter contribution to the cosmological constant based on a standard electroweak symmetry breaking scenario, during which the Higgs field develops a non-zero vacuum expectation value. Using equation~(\ref{vacuumenergy5}) rather than equation~(\ref{vacuumenergy3}), we expect to arrive at a lower theoretical value of the cosmological constant. Before electroweak symmetry breaking, the standard model particles are massless and do not yield any contribution to the renormalized vacuum energy. Thermal masses do not contribute to the cosmological constant, as they redshift with the scale factor and are therefore time-dependent. For simplicity, we only take the heaviest standard model particles into account after symmetry breaking: the top quark (\mbox{$m_{\mathrm{t}} = 173 \, \mathrm{GeV}$} and $n_{\mathrm{t}} = 12$ degrees of freedom), the Z boson (\mbox{$m_{\mathrm{Z}} = 91 \, \mathrm{GeV}$} and $n_{\mathrm{Z}} = 3$), and the W boson (\mbox{$m_{\mathrm{W}} = 80 \, \mathrm{GeV}$} and $n_{\mathrm{W}} = 6$). The top quark is a fermion and hence it contributes negatively to the energy density of the vacuum. Let us also assume that the Higgs boson has an approximate mass \mbox{$m_{\mathrm{H}} \simeq 150 \, \mathrm{GeV}$.} The Higgs potential has the form \mbox{$V(\phi) = - \frac{1}{2}\bar{m}^2 \phi^2 + \frac{1}{4}\lambda \phi^4$,} such that \mbox{$\bar{m}^2>0$} by construction. From \mbox{$\langle \hat{\phi}\rangle = v = 246\, \mathrm{GeV}$,} we find \mbox{$\bar{m} = m_{\mathrm{H}}/\sqrt{2} \simeq 106 \, \mathrm{GeV}$} and \mbox{$\lambda = \bar{m}^2/v^2 \simeq 0.186$.} Hence, the change of the Higgs potential during electroweak symmetry breaking follows roughly as \mbox{$\Delta V = - \frac{1}{4}\bar{m}^4/\lambda \simeq - 1.70 \cdot 10^8 \, \mathrm{GeV}^4$}. Finally, we have to make an assumption for the value of $\mu$, the renormalisation scale in equation~(\ref{vacuumenergy5}). We determine the value of the cosmological constant from measurements of photons originating from supernovae, whose wavelength is about \mbox{$\lambda_{\gamma} \simeq 500\,\mathrm{nm}$}. These photons couple to the metric, which in turn depends on the cosmological constant. Hence, the relevant renormalisation scale to consider is the mean of the photon's and graviton's energy, which roughly equals \mbox{$\mu \sim \sqrt{s} \sim \sqrt{E_{\gamma} E_{\mathrm{grav}}} \simeq 3\cdot 10^{-25} \,\mathrm{GeV}$.} We use \mbox{$E_{\mathrm{grav}} \simeq H_0 \simeq 3.7\cdot 10^{-41}\,\mathrm{GeV}$}, where we take \mbox{$H_0 \simeq 71\, \mathrm{km}/(\mathrm{s}\,\mathrm{Mpc})$} as the current Hubble parameter. Adding the various contributions, we find the following value of the change of the total energy density of the vacuum before and after electroweak symmetry breaking:
\begin{align}\label{estimatevacuumenergydensity}
\Delta \left [\langle 0| \hat{\rho}_{\mathrm{vac}} |0 \rangle \right] & = \sum_{i} \pm n_i \frac{m^4_i}{64 \pi^2}\log\left[\frac{m^2_i}{\mu^2}\right] + \Delta V \nonumber \\
& \simeq - 2 \cdot 10^{9} \, \mathrm{GeV}^{4}\,,
\end{align}
where the $+$ and $-$ apply for bosons and fermions, respectively. The geometric contribution to the cosmological constant $\Lambda$, which also contains the unknown, constant value of the potential energy of the Higgs field before electroweak symmetry breaking, thus has to be fine-tuned by 56 decimal places, i.e.: to a much lesser extent than in equation~(\ref{vacuumenergy3}). Although this amount of fine-tuning is clearly still unacceptable, it is, however, much better than the 123 orders of magnitude frequently quoted in literature. In order to further reduce the theoretically expected value of the cosmological constant, one would need to identify another, possibly unknown, symmetry principle, or find a mechanism which dynamically relaxes the cosmological constant to zero.

Often, global supersymmetry is invoked as the symmetry that can reduce the theoretical value of the cosmological constant. We point out that the estimate~(\ref{estimatevacuumenergydensity}) is lower than the contribution that would be produced by a broken supersymmetry. As supersymmetric partners of standard model particles have a larger mass than the particular standard model particle itself, broken supersymmetry would yield a contribution of the order of $\mathrm{TeV}^4$ or larger. Moreover, global supersymmetry is a hypothetical symmetry, still to be found in nature, while Lorentz symmetry is realised (to a high accuracy) in nature. Since furthermore gravity is essential in any consideration of the cosmological constant problem, local supersymmetry, in which the graviton acquires a supersymmetric partner too, is perhaps more natural than global supersymmetry. Unlike global supersymmetry, local supersymmetry does not automatically lead to a vanishing energy density of the vacuum.

The idea that relativistic invariance or Lorentz symmetry could play an important role when addressing the cosmological constant problem has been previously formulated by Akhmedov in \cite{Akhmedov:2002ts}, and has been subsequently expanded on by \cite{Ossola:2003ku}. We do, however, completely differ in the final results. Using cutoff renormalisation and Pauli-Villars regularisation, Akhmedov argues that the vacuum energy diverges quadratically with the cutoff scale, whereas we find a logarithmic dependence on the renormalisation scale. The reason is that he keeps a quadratic, divergent dependence on the cutoff, or, equivalently, on the Pauli-Villars mass, which he does \emph{not} remove using standard renormalisation techniques even though that term breaks Lorentz symmetry, as can clearly be seen from equations (\ref{vacuumenergy2}) and (\ref{vacuumenergy4}). Using Pauli-Villars renormalisation, one introduces a new mass scale to renormalise an expectation value. In order to decouple this new degree of freedom, one sends the Pauli-Villars mass to infinity to arrive at the final result. For completeness, we include this calculation in appendix B below.

Lorentz symmetry only reduces the matter part of the contribution to the cosmological constant. In this setup, the geometric contribution $\Lambda$ to $\Lambda_{\mathrm{eff}}$ is still a freely adjustable parameter in the theory. It would be intriguing to identify a principle, based on which \emph{both} the geometric contribution \emph{and} the matter contribution to $\Lambda_{\mathrm{eff}}$ can simultaneously be reduced. Such a scheme has recently been proposed by 't Hooft \cite{'tHooft:2011we}.

\textsc{Conclusion} --- We show that, by assuming a Lorentz invariant vacuum state in Minkowski spacetime, the theoretical value of the cosmological constant gets significantly reduced. Although this still does not solve the cosmological constant problem, a simple one loop estimate of the electroweak symmetry breaking transition indicates a fine-tuning of about 56 orders of magnitude of $\Lambda$ to account for the observed value of the vacuum energy density in the Universe. Moreover, Lorentz symmetry stabilises the gravitational hierarchy, in the sense that the cosmological constant, brought in accordance with Lorentz symmetry, runs logarithmically with scale. This means that, if one fixes the cosmological constant to the value observed today, $\rho_{\mathrm{eff}} \sim 10^{-47} \, \mathrm{GeV}^4$,
it will remain small (with respect to the quartic power of the running scale) when measured on a higher scale. As a final remark, we point out that massless particles such as the photon and the graviton do not yield any contribution to the cosmological constant at one loop level.

\textsc{Acknowledgements} --- We thank Gerard 't Hooft, Kimmo Kainulainen, and Willem Westra for useful comments and suggestions.

\

\textsc{Appendix A: Dimensional Renormalisation} --- Using dimensional regularisation, we are interested in evaluating:
\begin{subequations}
\label{vacuumenergy6}
\begin{align}
& \langle 0| \hat{\rho}_{\mathrm{vac}} |0 \rangle =
\int \frac{\mathrm{d}^{{\scriptscriptstyle{D}}-1}\vec{k}}{(2\pi)^{{\scriptscriptstyle{D}}-1}} \frac{1}{2} \sqrt{k^2+m^2} \label{vacuumenergy6a} \\
&\langle 0| \hat{P}_{\mathrm{vac}} |0 \rangle = \frac{1}{D-1} \int \frac{\mathrm{d}^{{\scriptscriptstyle{D}}-1}\vec{k}}{(2\pi)^{{\scriptscriptstyle{D}}-1}} \frac{k^2}{ \sqrt{k^2+m^2}} \label{vacuumenergy6b} \,.
\end{align}
\end{subequations}
These integrals can be evaluated in a straightforward manner in a dimension where they converge. Analytical continuation yields:
\begin{equation}\label{vacuumenergy7}
\langle 0| \hat{T}_{\mu\nu} |0 \rangle  = \frac{m^{\scriptscriptstyle{D}}\, \Gamma \left( - \frac{D}{2}\right)}{2^{{\scriptscriptstyle{D}}+1} \pi^{\frac{D}{2}}} g_{\mu\nu} \,.
\end{equation}
This is clearly a Lorentz invariant result under \mbox{$SO(D-1,1)$.} By noting that:
\begin{equation}\label{vacuumenergy8}
\Gamma \left( - \frac{D}{2}\right) = - \frac{8 \Gamma \left( 3 - \frac{D}{2}\right)}{(D-4)(D-2)D} \simeq \frac{-1}{D-4} + \frac{3-2\gamma_{\mathrm{E}}}{4}\,,
\end{equation}
one can isolate the divergence in $D=4$ in a straightforward manner (see e.g. \cite{Peskin:1995ev, Koksma:2009wa,Koksma:2011fx} for a more elaborate use of dimensional regularisation). Note that the authors of \cite{Ossola:2003ku} have separately considered the pole in $D=2$, which is not in accordance with standard dimensional regularisation procedures. The local, Lorentz invariant counterterm thus follows as:
\begin{equation}\label{vacuumenergy9}
T_{\mu\nu}^{\mathrm{ct}}  = \frac{m^{4} \mu^{{\scriptscriptstyle{D}}-4}}{2^5 \pi^2} \frac{1}{D-4} g_{\mu\nu}\,,
\end{equation}
where we introduce the renormalisation scale $\mu$ again. The renormalised stress-energy tensor in $D=4$ now reads:
\begin{equation}\label{vacuumenergy10}
T_{\mu\nu}^{\mathrm{ren}}  = - \frac{m^4}{64\pi^2}\left[ \log \left( \frac{m^2}{4 \pi\mu^2} \right) - \frac{3}{2} + \gamma_{\mathrm{E}} \right]  g_{\mu\nu}\,.
\end{equation}
Here, we use a minimal subtraction scheme to remove the divergences. If we were to use a non-minimal scheme, we would find complete agreement with equation (\ref{vacuumenergy5}).

\textsc{Appendix B: Pauli-Villars Renormalisation} --- For completeness, let us check that we also confirm the result (\ref{vacuumenergy5}) by making use of the Pauli-Villars regularisation method. In order to remove the divergences in the stress-energy tensor, one introduces four heavy fictitious scalar fields, with masses $M_{i}$, where $i=\{1,2,3,4\}$, with a corresponding stress-energy tensor that reads:
\begin{equation}\label{vacuumenergy11}
T_{\mu\nu}^{\mathrm{PV}}  = \sum_{i=1}^{4} c_{i} \int \frac{\mathrm{d}^{3}\vec{k}}{(2\pi)^{3}} \frac{k_{\mu}k_{\nu}}{2 \sqrt{k^2+M_i^2}}\,,
\end{equation}
where we choose the coefficients $c_i$ such that the divergences in equation (\ref{vacuumenergy2}) are regulated. Imposing the following four conditions on the $c_{i}$:
\begin{subequations}
\label{vacuumenergy12}
\begin{align}
\sum_{i=1}^4 c_i =& -1 \label{vacuumenergy12a}\\
\sum_{i=1}^4 c_i M_i^2 =&  -m^2 \label{vacuumenergy12b}\\
\sum_{i=1}^4 c_i M_i^4 =& -m^4 \label{vacuumenergy12c}\\
\sum_{i=1}^4 c_i M_i^4 \log \left(\frac{M_i^2}{\mu^2}\right)=& \,\, 0\label{vacuumenergy12d}\,,
\end{align}
\end{subequations}
confirms the renormalised and Lorentz invariant stress-energy tensor (\ref{vacuumenergy5}). As before, we introduce a renormalisation scale $\mu$. We can now first send $\Lambda \rightarrow \infty$ to remove the dependence on the cutoff. Then, we send the $M_i \rightarrow \infty$ to decouple the fictitious Pauli-Villars fields. We emphasise that, when one imposes Lorentz invariance, the final result for the renormalised vacuum stress-energy tensor does not depend on the regularisation procedure. We again disagree with the arguments presented in \cite{Akhmedov:2002ts, Ossola:2003ku}.

\end{document}